\documentclass{elsart}

\usepackage{natbib}

 \usepackage{graphics}
 \usepackage{graphicx}
 \usepackage{epsfig}

\usepackage{amssymb}

 \usepackage{lineno}
 \usepackage{amsmath}
\usepackage{amssymb}
\usepackage{pifont}
\usepackage{longtable,lscape}
\usepackage[varg]{txfonts}
\usepackage{longtable}
\usepackage{graphicx}
\usepackage{txfonts}

\begin{document}

\begin{frontmatter}

\title{Seismic analysis of 70 Ophiuchi A: A new quantity proposed}


\author[1,3]{Y. K. Tang},
\author[2,1]{S. L. Bi}
and
\author[1,3]{N. Gai}

\address[1]{National Astronomical Observatories/Yunnan Observatory, Chinese Academy of Sciences, Kunming 650011,
   P. R. China\\ e-mail: bisl@bnu.edu.cn; tangyanke@ynao.ac.cn}

\address[2]{Department of Astronomy Beijing Normal University, Beijing 100875,
  P. R. China}

\address[3]{Graduate School of the Chinese Academy of Sciences, Beijing 100039, P. R. China}
\begin{abstract}
  The basic intent of this paper is to model 70 Ophiuchi A using the
latest asteroseismic observations as complementary constraints and
to determine the fundamental parameters of the star. Additionally,
we propose a new quantity to lift the degeneracy between the initial
chemical composition and stellar age. Using the Yale stellar
evolution code (YREC7), we construct a series of stellar
evolutionary tracks for the mass range $M$ = 0.85 -- 0.93
  $M_{\odot}$ with different composition $Y_{i}$ (0.26 -- 0.30) and $Z_{i}$ (0.017 -- 0.023). Along these tracks, we
   select a grid of stellar model candidates that fall within the error box in
   the HR diagram to calculate the theoretical frequencies, the large- and
   small- frequency separations using the Guenther's stellar pulsation
   code. Following the asymptotic formula of stellar $p$-modes, we define a quantity $r_{01}$ which is correlated with stellar
   age. Also, we test it by theoretical adiabatic frequencies of many models.
Many detailed models of 70 Ophiuchi A have been listed in Table 3.
By combining all non-asteroseismic observations available for 70
Ophiuchi A with these seismological data, we think that Model 60,
Model 125 and Model 126, listed in Table 3, are the optimum models
presently. Meanwhile, we predict that the radius of this star is
about 0.860 -- 0.865 $R_{\odot}$ and the age is about 6.8 -- 7.0 Gyr
with mass 0.89 -- 0.90 $M_{\odot}$. Additionally, we prove that the
new quantity $r_{01}$ can be a useful indicator of stellar age.
\end{abstract}


\end{frontmatter}

\section{Introduction}
\label{}

The solar five-minute oscillations have led to a wealth of
information about the internal structure of the Sun. These results
stimulated various attempts to detect solar-like oscillations for a
handful of solar-type stars. Individual $p$-mode frequencies have
been identified for a few stars: $\alpha$ Cen A (Bouchy and Carrier,
2002; Bedding et al., 2004), $\alpha$ Cen B (Carrier and Bourban,
2003a; Kjeldsen et al., 2005), $\mu$ Arae (Bouchy et al., 2005), HD
49933 (Mosser et al., 2005), $\beta$ Vir (Marti\'{c} et al., 2004a;
Carrier et al., 2005b), Procyon A (Marti\'{c} et al., 2004b;
Eggenberger et al., 2004a), $\eta$ Bootis (Kjeldsen et al., 2003;
Carrier et al., 2005a), $\beta$ Hyi (Bedding et al., 2001; Carrier
et al., 2001) and $\delta$ Eri (Carrier et al., 2003b). Based on
these asteroseismic data, numerous theoretical analyses have been
performed in order to determine precise global stellar parameters
and to test the various complicating physical effects on the stellar
structure and evolutionary theory (Th\'{e}venin et al., 2002;
Eggenberger et al., 2004b, 2005; Kervella et al., 2004; Miglio and
Montalb\'{a}n, 2005; Provost et al., 2004, 2006).\\
                                                        \\
 Recently, Carrier and Eggenberger (2006) detected solar-like oscillations on
the K0 V star 70 Ophiuchi A (HD 165341), and identified some
possible existing frequencies. They obtained the large separation
$\Delta \nu=161.7\pm0.3$ $\mu Hz$ by observation over 6 nights with
HARPS. The spectroscopic visual binary system 70 Ophiuchi is one of
our nearest neighbors (5 pc) and is among the first discovered
binary stars. It was observed first by Herschel in 1779.  So 70
Ophiuchi A is famous as the primary of a visual and spectroscopic
binary system in the solar neighborhood. Although many observation
data have been obtained since 1779, the theoretical analysis of 70
Ophiuchi A has only been made  by Fernandes et al. (1998). By a
calibration method which take into account the simultaneous
evolution of the two members of the binary system, they analyzed the
70 Ophiuchi A by means of standard evolutionary stellar models using
the CESAM code bf (Morel, 1997) without microscopic diffusion. They
found that the metallicity of 70 Ophiuchi A is very close to the
solar one, the values of mixing-length parameter $\alpha$ and helium
abundance Y are near the Sun. They thought that the star is younger
than the Sun and $3\pm 2$ Gyr is probably an limit considering the
age versus stellar rotation relation with its rotation velocity
($vsini\approx16$ $km$ $s^{-1}$).\\
                                   \\
The aim of our paper is to present the model which can be
constrained by these seismology data. The observational constraints
available for 70 Ophiuchi A are summarized in Sect. 2, while the
numerical calculations are presented in Sect. 3. The seismic
analyses are carried out and a new quantity $r_{01}$ as a indication
of stellar age is proposed in Sect. 4. Finally, the discussion and
conclusions are given in Sect. 5.

\section{Observational Constraints}

\subsection{Non-asteroseismic observation constraints }

\begin{table}
\caption{Non-asteroseismic observational data of 70 Ophiuchi A. }
\begin{tabular}{c c c}
\hline\hline
Observable & Value & Source \\

 \hline
 Mass $M/M_{\odot}$ & $0.89\pm0.04$  & (1)   \\
 \hline
 Effective temperature $T_{eff}$(K) & $5322\pm20$  & (2)   \\
\hline
Luminosity $\textrm{log}(L/L_{\odot})$&$-0.29\pm0.03$& (1)\\
\hline

Metallicity $[Fe/H]_{s}$&$0.0\pm0.1$& (1)\\
\hline

Surface heavy element abundance$[Z/X]_{s}$&$0.02365\pm0.00535$& (3)\\

 \hline

 \end{tabular}\\
  References.---(1) Fernandes et al. (1998), (2) Gray and Johnson (1991), (3) this
  paper.
  \end{table}
The mass of this star was investigated by Batten et al. (1984),
Heintz et al. (1988), Fernandes et al. (1998) and Pourbaix et al.
(2000), respectively. In the paper, we adopt the value of mass
deduced from Fernandes et al. (1998). The effective temperature was
determined by Gray and Johnson (1991). So far, the metallicity
obtained by observation are [Fe/H] = -0.05 (Peterson, 1978) and
[Fe/H] = 0.00 (Perrin et al., 1975). We choose the [Fe/H] = 0.0
$\pm$ 0.1 as a representative value according to Fernandes  et al.
(1998).

The mass fraction of heavy elements, $Z$, was derived assuming
$\textrm{log}[Z/X] \approx [Fe/H] + \textrm{log} [Z/X]_{\odot}$  and
$[Z/X]_{\odot} = 0.0230$ (Grevesse and Sauval, 1998), for the solar
mixture. So we can deduce the $[Z/X]_{s}$ = 0.0183 -- 0.0290.

All non-asteroseismic observational constraints are listed in Table
1.
\subsection{Asteroseismic constraints}

 Solar-like oscillations of 70 Ophiuchi A have been detected by
Carrier and Eggenberger (2006) with the HARPS spectrograph. Fourteen
individual modes are identified with amplitudes in the range 11 to
14 $cm$ $s^{-1}$. Although they listed two groups of frequencies by
mode identification (see Table 2 in Carrier and Eggenberger, 2006),
one group of frequencies with a average large separation
$\Delta\nu=161.7$ $\mu Hz$ was suggested to be more reliable than
the other with a average large separation $\Delta\nu=172.2$ $\mu
Hz$. The star 70 Ophiuchi A is very similar to $\alpha$ Cen B with
the same spectral type and similar large separation, which has a
mean small separation of 10 $\mu Hz$. It is thought that the small
separation should be similar. By inspecting the results of the mode
identification, they note that the value of the small separation
coming from the identification with the large separation of 172.2
$\mu Hz$ is significantly different from 10 $\mu Hz$. If the large
separation is 172.2 $\mu Hz$, the small separation will be lower
than 6.5 $\mu Hz$ in the frequency range 3 -- 4.5 $mHz$. Although
this identification is less reliable than the one with a large
separation of 161.7 $\mu Hz$, the solution $\Delta \nu$ = 172.2 $\mu
Hz$ can not be ruled out definitely. We refer to these two groups of
results in the paper and make analyses in Sect. 4 and Sect. 5.
%
%

\section{Stellar models}

\begin{table}
\caption{Input parameters for model tracks.}
\begin{tabular}{c c c c}
\hline\hline
Variable & Minimum Value & Maximum Value & $\delta$ \\
 \hline
 Mass $M/M_{\odot}$ & 0.85 & 0.93 & 0.01   \\
\hline
 Initial heavy element abundance $Z_{i}$ &0.017&  0.023 & 0.001\\
\hline

Initial Helium abundance $Y_{i}$&$0.26$& 0.30 & 0.01\\

 \hline

 \end{tabular}\\
 Note.---The value $\delta$ defines the increment between minimum
 and maximum parameter values used to create the model array.
   \end{table}

\begin{figure*}
\includegraphics[angle=0,width=7cm,height=7cm]{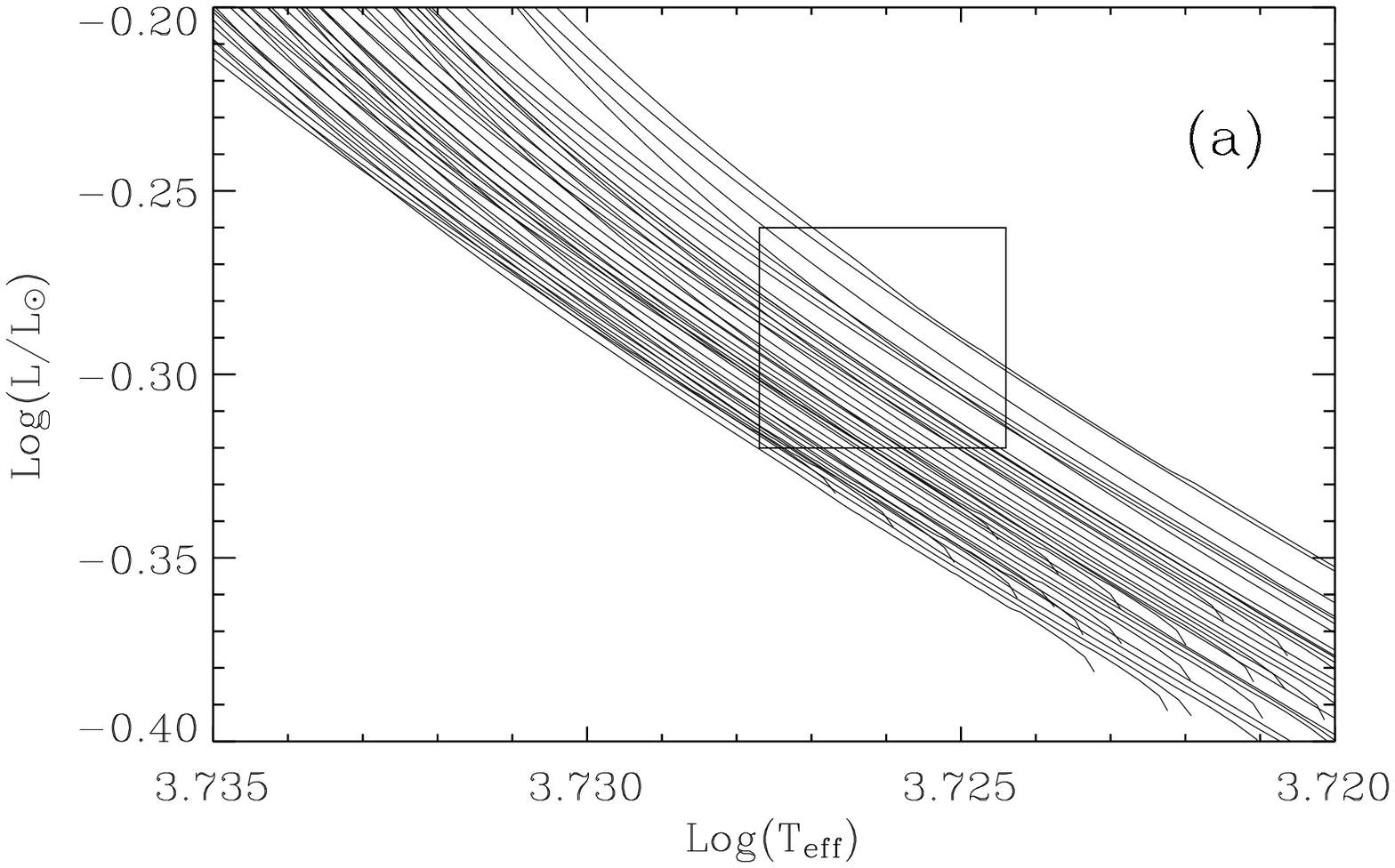}%
\includegraphics[angle=0,width=7cm,height=7cm]{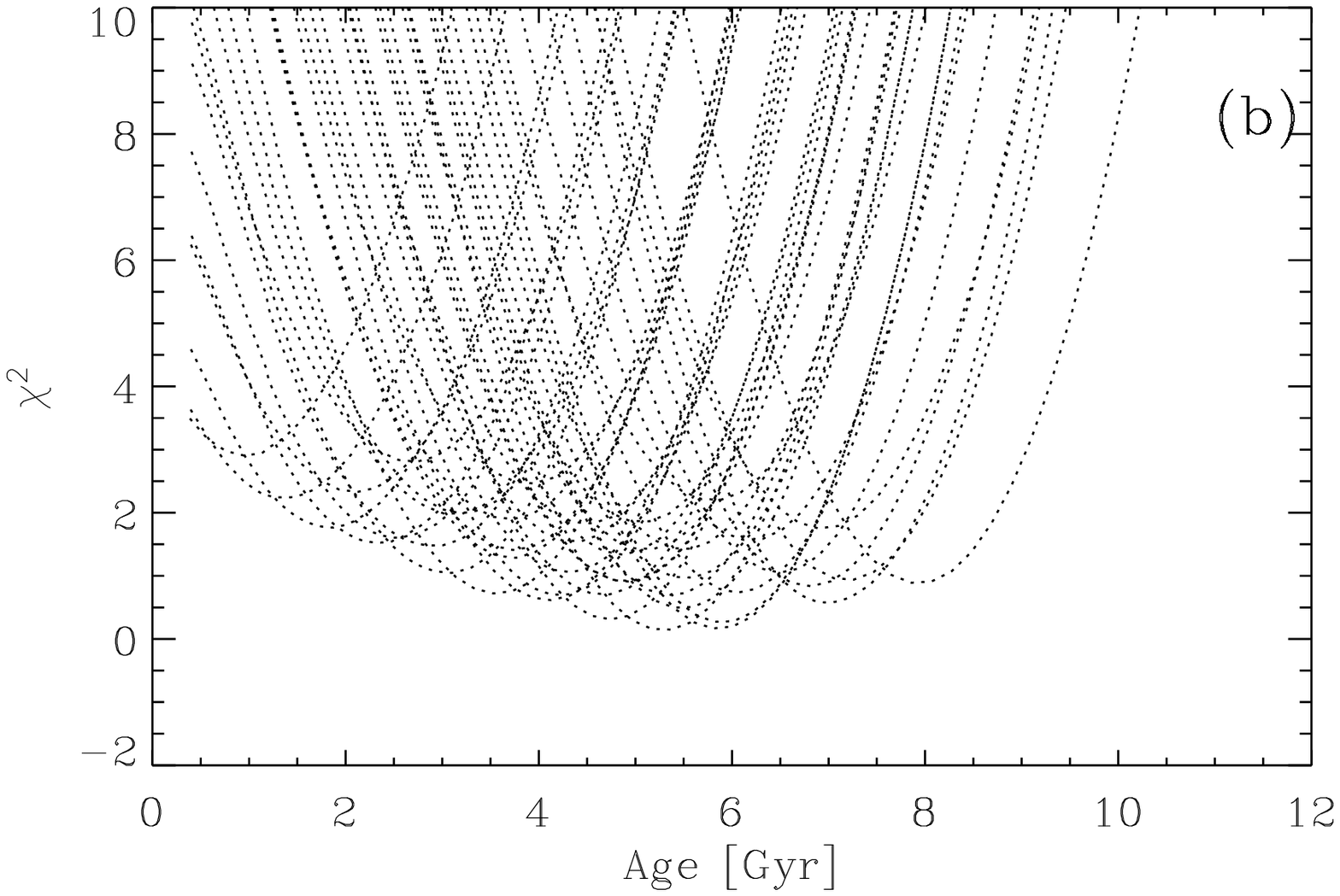}
\caption{Evolutionary tracks in the HR diagram and $\chi^{2}$ as a
function of age for 70 Ophiuchi A. a) The selected individual
stellar evolutionary tracks (44 in total). b) $\chi^{2}$ values
calculated for 70 Ophiuchi A observational data using different
$Z_{i}$, $Y_{i}$ and Mass, plotted as a function of age. $\chi^{2}$
refers to non-asteroseismological observables as denoted by eq. (1).
}
\end{figure*}

 We will construct a grid of stellar evolutionary models by Yale stellar
evolution code (YREC; Guenther et al., 1992) with microscopic
diffusion. The initial zero-age main sequence (ZAMS) model
 used for 70 Ophiuchi A was created from pre-main sequence evolution calculations.
 In these computations, we do not consider
 rotation and magnetic field effect. These models are computed using OPAL equation of state
tables EOS2001 (Rogers and Nayfonov, 2002), the opacities
interpolated between OPAL GN93 (Iglesias and Rogers, 1996) and low
temperature tables (Alexander and Ferguson, 1994). Using the
standard mixing-length theory, we set $\alpha=1.7$ for all models,
close to the value which is required to reproduce the solar radius
under the same physical assumptions and stellar evolution code.
Meanwhile, it must be emphasized that there are still a number of
uncertainties in our analyses, foremost among which is the still
open question of mixing-length theory responsible for the stellar
model. The nuclear reaction rates have been updated according to
Bahcall and Pinsonneault (1995). The Krishna-Swamy Atmosphere
T-$\tau$ relation is used for this solar-like star (Guenther and
Demarque, 2000). Also, we consider the microscopic diffusion effect,
by using the diffusion coefficients of Thoul et al. (1994). Since 70
Ophiuchi A, like $\alpha$ Cen B, is less massive than the Sun, the
mass contained in its convective zone is much larger and, therefore,
the effect of microscopic diffusion is much smaller (Miglio and
Montalb\'{a}n, 2005; Morel and Baglin, 1999). However, it is
necessary to consider this effect as physical process in stellar
modeling (see Provost et al., 2005, 2006).

\begin{figure}
\includegraphics[angle=0,width=14cm,height=12cm]{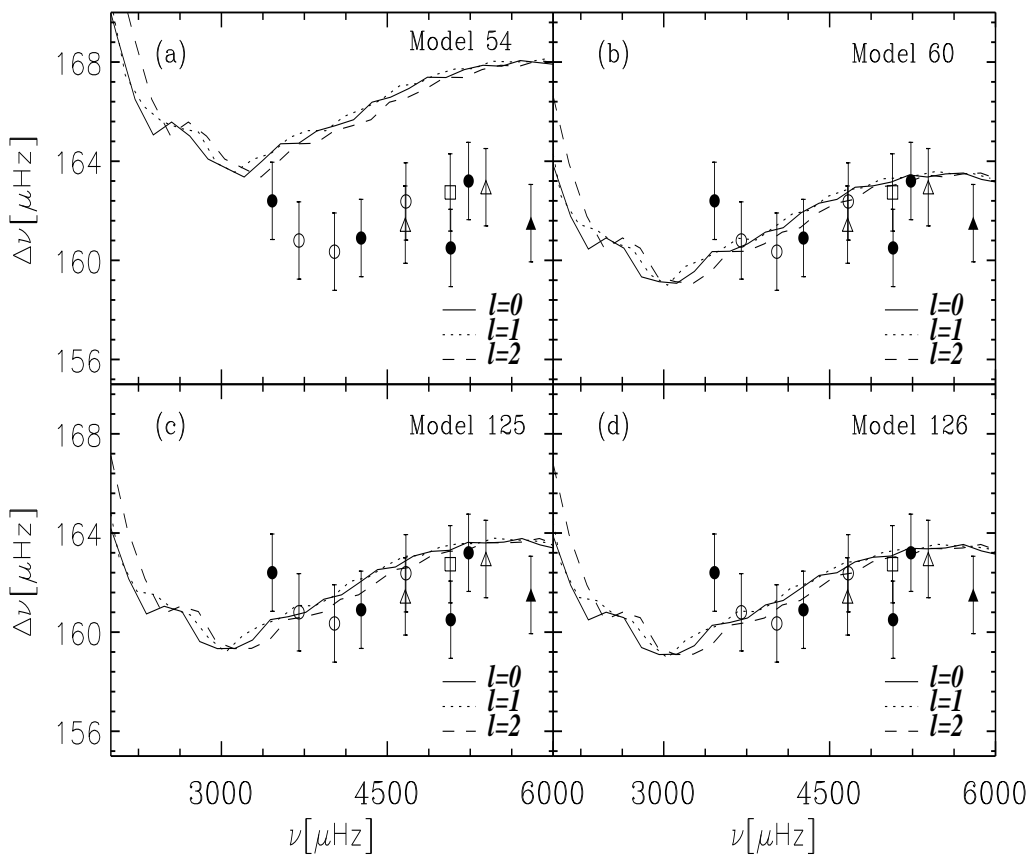}
\caption{Large-frequency separations vs. frequency for the Models
54, 60, 125, 126 (in Table 3). The observable large separations
$\Delta \nu$ versus frequency for $p$-modes of degree
$l=0$($\bullet$), $l=1$($\blacktriangle$)and $l=2$($\blacksquare$)
are obtained from Carrier and Eggenberger (2006), which correspond
to the average large separation of 161.7 $\mu Hz$. Open symbols
correspond to large separation averages taken between non-successive
modes and vice versa. All individual errors are fixed to
$\sqrt{2}\times 1.1$ $\mu Hz$ (half resolution).}
\end{figure}

In general, the determination of parameters $(M,t, Y_{i}, Z_{i})$
fitting the observational constraints needs two steps. The first
step is to construct a grid of models with position in the HR
diagram in agreement with the observational values of the
luminosity, the effective temperature and the surface metallicity.
The principal constraints deduced from non-asteroseismic observation
are listed in Table 1. The error box, which is composed of
observational effective temperature and luminosity, represents the
possible position of 70 Ophiuchi A in the HR diagram (see Fig. 1a).
According to the results of Fernandes et al. (1998), we list the
parameter space of mass M, the initial heavy-element abundance
$Z_{i}$  and the initial helium abundance $Y_{i}$ in Table 2. Since
the microscopic diffusion is included in our paper, we give the
wider parameter space of initial heavy-element abundance $Z_{i}$
than the range of $Z_{i}$ of Fernandes et al. (1998).

By adjusting three parameters $M$, $Y_{i}$ and $Z_{i}$ listed in
Table 2, we can obtain many evolutionary tracks passing through the
error box in the HR diagram. Now we consider a function which
describes the agreement between the observations and the theoretical
results:

\begin {equation}
\chi^{2}\equiv\sum^{3}_{i=1}(\frac{C^{theo}_{i}-C^{obs}_{i}}{\sigma
C^{obs}_{i}})^{2},
\end {equation}
where $\mathbf{C}$ represent the following quantities:
$L/L_{\odot}$, $T_{eff}$ and $[Z/X]_{s}$, $\mathbf{C}^{theo}$
represent the theoretical values and $\mathbf{C}^{obs}$ represent
the observational values listed in Table 1. The vector $\sigma
\mathbf{C}_{i}^{obs}$ contains the errors on these observations
which are also given in Table 1.

As Fernandes(1998) has pointed that the age of 70 Ophiuchi A is
$3\pm 2$ Gyr, it is reasonable for us to choose the evolutionary
tracks passing through the error box within 8 Gyr. We select 44
evolutionary tracks passing through error box as our possible
candidates to go on with our investigations. Fig. 1a gives 44
evolutionary tracks, and Fig. 1b presents $\chi^{2}$ as a function
of the age correspondingly. It is well-known that $\chi^{2}$ is
smaller, the more competitive is the candidate. Fig. 1b shows that
models with $\chi^{2}$ smaller than 1 have ages between 3Gyr to
7Gyr. From Fig. 1a, we find that the upper section of error box is
empty.  The reason of the empty upper section of the error box is
related to the range of initial parameters, like mass, initial
composition and specially the mixing length parameter. We think that
the future interferometric measurement of the radius could reduce
the domain of the possible position in the HR diagram (e.g., Provost
et al., 2006).

The second step is to determine the optimum model using the
asteroseismic measurements. We will select a grid of models along
these 44 tracks shown in Fig. 1a to calculate the low-$l$ $p$-modes
frequencies. We list the representative models extracted from every
tracks in Table 3.

The detailed pulsation analysis is described in the next section.

\section{Pulsation analysis}
\subsection{Selecting the optimum model}

\begin{figure}
\includegraphics[angle=0,width=14cm,height=12cm]{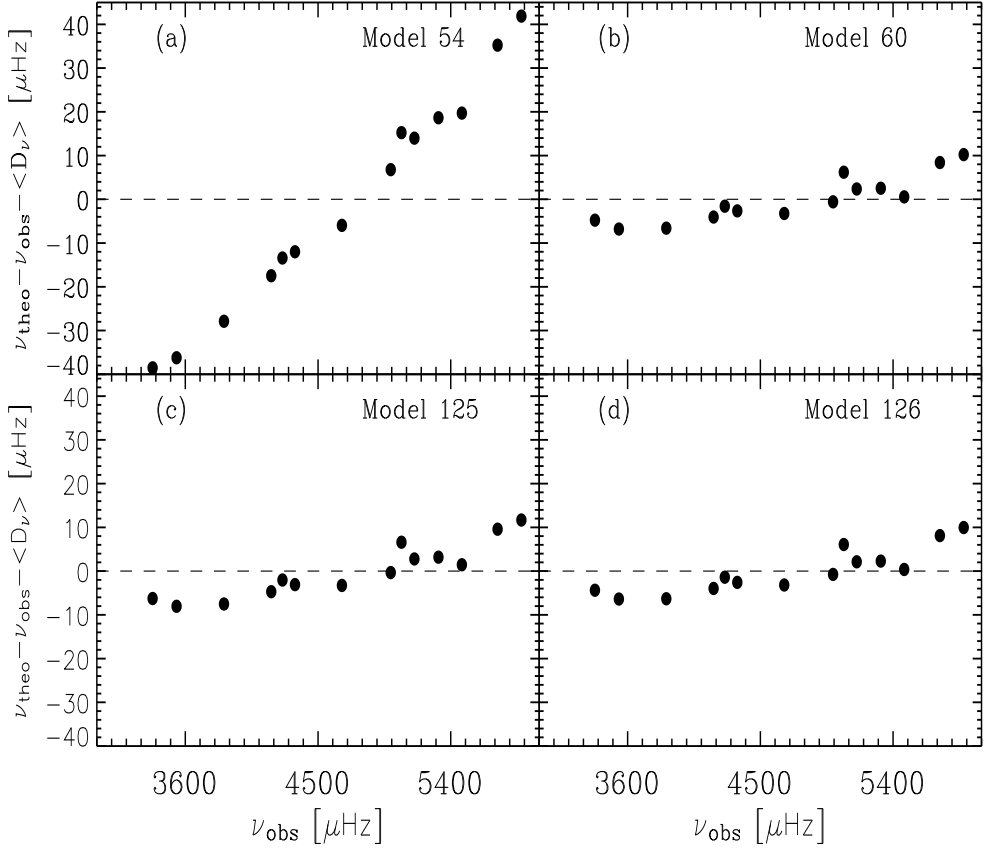}
\caption{Differences between calculated and observable frequencies
for the Model 54, 60, 125, 126 in Table 3. The systematic shifts
$\langle D_{\nu}\rangle$ for the four models are 30.059, 63.75,
70.252, 63.24 $\mu Hz$, respectively .  The observable frequencies
correspond to the average large separation of 161.7 $\mu Hz$ (see
text for details.}
\end{figure}

\begin{figure}
\includegraphics[angle=0,width=14cm,height=12cm]{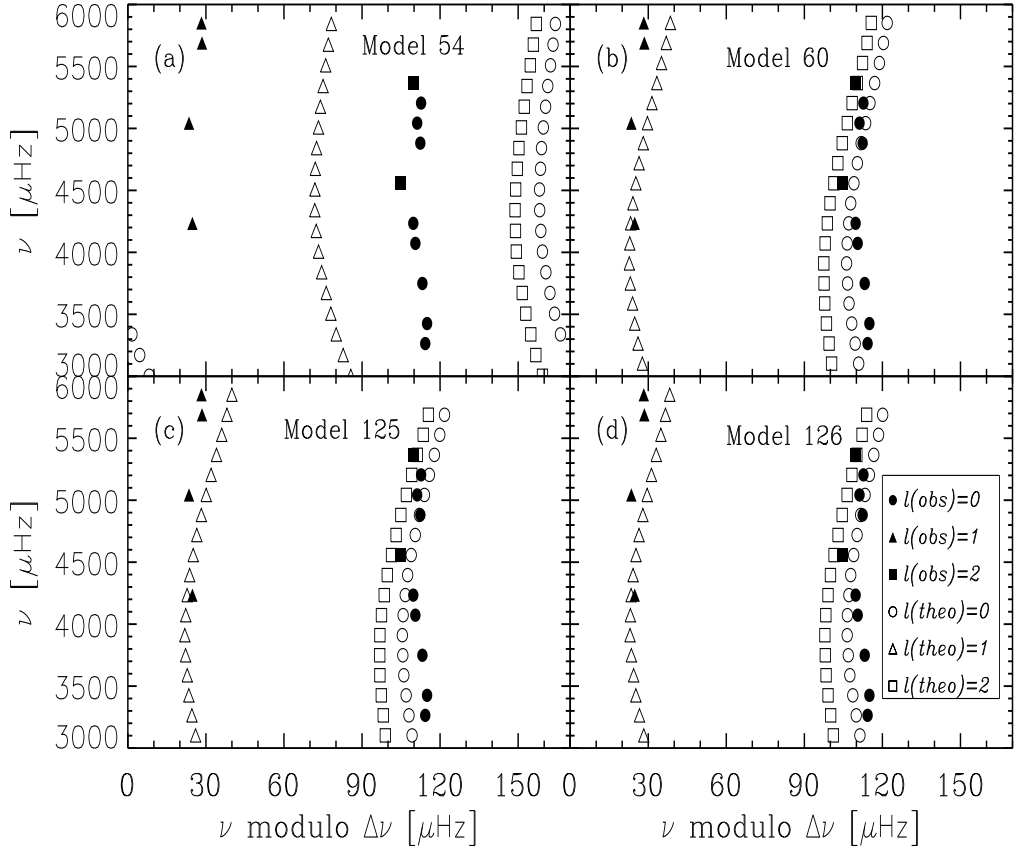}
\caption{Echelle diagrams for the Models 54, 60, 125, 126 (in Table
3), which average large separations $\langle\Delta\nu\rangle$ are
166.22, 161.7, 161.92, 161.68 $\mu Hz$, respectively.   The
systematic shift $\langle D_{\nu}\rangle$, which have been applied
to the theoretical frequencies, are 30.059, 63.75, 70.252, 63.24
$\mu Hz$, for the four models respectively.  Open symbols refer to
the theoretical frequencies, and filled symbols to the observable
frequencies. Circles are used for $l$ = 0 modes, triangles for $l$ =
1 modes, and squares for $l$ = 2 modes. The observable frequencies
correspond to the average large separation of 161.7 $\mu Hz$ (see
text for details).}
\end{figure}

Using Guenther's pulsation code (Guenther, 1994), we calculate the
 adiabatic low-$l$ $p$-mode frequencies of the selected models.
 We define the
large separations $\Delta\nu$ and small separations $\delta\nu$ in
the usual way (Tassoul, 1980):
\begin {equation}
\Delta \nu_{n,l}\equiv\nu_{n,l}-\nu_{n-1,l}
\end {equation}
and
\begin {equation}
\delta \nu_{n,l}\equiv\nu_{n,l}-\nu_{n-1,l+2} ,
\end {equation}
where $n$ is the radial order, $l$ is the degree, and $\nu$ is the
frequency. Because the expected acoustic cutoff has a limit, we only
calculate the mean large- and small- separations by averaging over
$n$ = 10 -- 30 (See Murphy et al., 2004). Within these 44 tracks, we
list 129 models in Table 3. $\langle\Delta\nu_{l}\rangle$ represents
the mean of large separations $\Delta\nu_{n,l}$ for $n$ = 10 to 30.
The frequency range corresponds to about 2000 $\mu HZ$--6000 $\mu
Hz$. Additionally, $\langle\Delta\nu\rangle$ represents the mean of
$\langle\Delta\nu_{l}\rangle$ for $l=0$ to 3. In the same way,
$\langle\delta\nu_{02}\rangle$ and $\langle\delta\nu_{13}\rangle$
represent the mean of $\delta\nu_{n,0}$ and $\delta\nu_{n,1}$ for
$n$ = 10 to 30, respectively. So far, we only know the large
separations and the fourteen individual modes of the star based on
the asteroseismic data of Carrier and Eggenberger (2006). Guenther
(1998) pointed that the large separations are most easily
identifiable characteristics in the $p$-mode spectrum. Because they
are seen as a peak in the Fourier transform of the power spectrum
and they are mostly uncontaminated by composition effects, these
large separations  provide an efficient way to constrain stellar
model. It is also important to remember that the theoretical
frequencies calculated in our paper should not be expected to match
the observed frequencies of Carrier and Eggenberger (2006)
perfectly. We think that there are three reasons. Firstly, our
theoretical models do not match the mass and radius of the star
precisely. Secondly, the uncertainty in calculating the sound speed
in the outer layers of the models comes into being, where
non-adiabatic effects become important. Thirdly, at high
frequencies, the effect of the convection-oscillation interactions
is larger and the description of convection is open problem.
Although the differences between the theoretical frequencies and the
observed frequencies could result in significant effect on the large
separations, we think that the effect is small due to the large
separations correspond to differences between frequencies of modes
with the same angular degree $l$ and consecutive radial order $n$.
Therefore, in our paper, we think that the matching the observable
large separations is the important criterion to select the optimum
model. In Table 3, we find that the average large separations of
Model 60, Model 125 and Model 126 are 161.7, 161.92 and 161.68 $\mu
Hz$, in good agreement with the mean value derived from Carrier and
Eggenberger (2006). So we can tentatively say that these models may
be the best fit models. In Fig. 2, we plot the observational results
about the large separations and the errors. Also we plot the large
separation as a function of frequency for Model 54 in Fig. 2a, Model
60 in Fig. 2b, Model 125 in Fig. 2c and Model 126 in Fig. 2d. We
clearly find that the theoretical large separations of the Model 60,
Model 125 and Model 126 are consistent with the observations. Model
54, as the representative of many non-fit models, is not consistent
with the observational large separations. Therefore, we have
sufficient reasons to say that Model 60 , Model 125 and Model 126
are really the best fit models. Meanwhile, we can predict that the
radius of star is 0.860 -- 0.865 $R_{\odot}$ and the age is about
6.8 -- 7.0 Gyr with mass 0.89 -- 0.90 $M_{\odot}$ presently.

Once the asteroseismic observation can confirm the large separations
to be $161.7 \pm 0.3$ $\mu Hz$ and the theory Model 60 , Model 125,
Model 126 are considered as the best models, we can predict that the
mean small separation $\langle\delta \nu_{02}\rangle$ is about 10.29
-- 10.48 $\mu Hz$ and the radius of the star is about 0.860 -- 0.865
$R_{\odot}$. Direct measurements of stellar diameters from
interferometric observations should provide an independent check for
asteroseismic predictions such as Kervella et al. (2003a, 2003b).

 In order to compare the
theoretical $p$-mode frequencies deduced from the models in Table 3
with the observational frequencies provided by Carrier and Eggenber
(2006), we plot the echelle diagram of every model and find that no
model can fit observational frequencies.
 For the exact values of the frequencies, considering above three reasons,
  a linear shift of a few $\mu Hz$ between theoretical and
observational frequencies is perfectly acceptable. Taking into
account it, we define the mean value of the difference between the
theoretical and observational frequencies (e.g., Eggenberger et al.,
2004b, 2005):

\begin {equation}
\langle D_{\nu}\rangle\equiv
\frac{1}{N}\sum_{i=1}^{N}(\nu_{i}^{theo}-\nu_{i}^{obs}).
\end {equation}
where $N$ is the number of observable frequencies ($N$ = 14).

 Takeing into account the systematic difference $\langle
D_{\nu}\rangle$ between theoretical and observable frequencies, we
plot the differences between calculated and observed frequencies in
Fig. 3 and the echelle diagram in Fig. 4. The observable frequencies
correspond to the average large separation of 161.7 $\mu Hz$ in
these figures. Fig. 3a, Fig. 3b, Fig. 3c and Fig. 3d correspond to
the Model 54 with $\langle D_{\nu}\rangle = 30.059$ $\mu Hz$, Model
60 with $\langle D_{\nu}\rangle = 63.75$ $\mu Hz$, Model 125 with
$\langle D_{\nu}\rangle = 70.252$ $\mu Hz$, and Model 126 with
$\langle D_{\nu}\rangle = 63.24$ $\mu Hz$, respectively. Fig. 4a,
Fig. 4b, Fig. 4c and Fig. 4d show the echelle diagram of the Model
54, Model 60, Model 125 and Model 126 respectively. For $p$-modes in
the asymptotic theory ($n \gg l$), the large separations are nearly
constant; meanwhile the so-called ``echelle diagrams" present the
frequencies in ordinates, and the same frequencies modulo the
average large separation in abscissa. So the asymptotic theory
predicts an approximated vertical line for given degree. In this
case, Fig. 4b, Fig. 4c and Fig. 4d show that the theoretical
frequencies of Model 60, Model 125, Model 126 can fit the observable
frequencies with 161.7 $\mu Hz$. Meanwhile, we find that the
systematic differences $\langle D_{\nu}\rangle$ are larger than the
results of $\alpha$ Cen B obtained by Eggenberger (2004b). It is
interesting to analyze the difference in future work.

\subsection{Asymptotic formula and frequency analysis}

\subsubsection{Large Separations and Small Separations}

\begin{figure}
\includegraphics[angle=0,width=9cm,height=7cm]{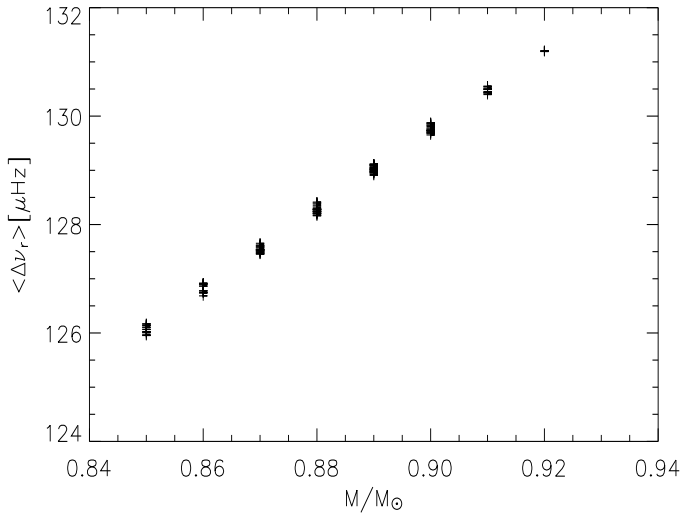}
\caption{The ``reduced" large separation vs. mass for each of the
129 stellar models.}
\end{figure}

\begin{figure}
\includegraphics[angle=0,width=9cm,height=7cm]{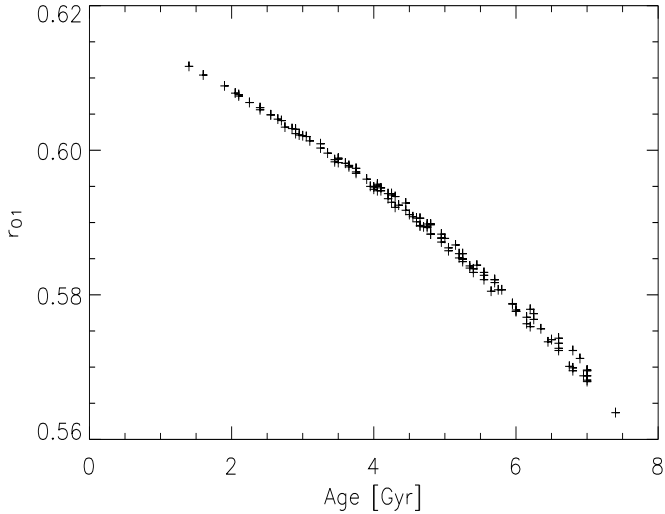}%
\caption{The ratio of small separations adjacent in $l$ vs. age for
each of 129 stellar models.}
\end{figure}

It is well known from asymptotic theory that the large separations
are mainly sensitive to the stellar radius (Tassoul, 1980;
Christensen-Dalsgaard, 1984). More precisely, the asymptotic
behavior of $\Delta\nu$ is expected to scale with $(M/R^{3})^{1/2}$,
where $M$ is the mass of the star and $R$ is its radius. Meanwhile,
Murphy et al. (2004) find that a degeneracy in predicted radius
occurs for models of different mass. Here, the degeneracy means that
the $\langle\Delta\nu\rangle$ changes with radius and mass (see Fig.
4 in Murphy et al., 2004). In order to lift the degeneracy,
Fernandes and Monteiro (2003) and Murphy et al. (2004) assumed
homology to compare theoretical models by introducing a ``reduced"
radius, such as
\begin {equation}
\langle\Delta\nu_{r}\rangle=\langle\Delta\nu_{n,l}\rangle(R/R_{\odot})^{3/2}.
\end {equation}
Here, we name the quantity $\langle\Delta\nu_{r}\rangle$ ``reduced"
large separation. We draw the ``reduced" large separation
$\langle\Delta\nu_{r}\rangle$ versus mass in Fig. 5 and list the
values of $\langle\Delta\nu_{r}\rangle$ for each model in Table 3.
From Fig. 5, we find that the degeneracy was lifted approximately.
It is easily seen that the values of the
$\langle\Delta\nu_{r}\rangle$ are relatively consistent with each
mass. It is successful using the $\langle\Delta\nu_{r}\rangle$
instead of large separations to indicate the stellar mass.

The small separations, like the large separations , will be visible
as peaks in the Fourier transform of the power spectrum. At the
earlier stage, Christensen-Dalsgaard (1984) proposed that the
calculation of small separations could put a constraint on the age
the star. Subsequently, Ulrich (1986) realized that only if the
composition of the star is known completely can one use the small
separations to correctly identify a stellar age. This point has been
illustrated in Murphy et al. (2004). Thus, the various chemical
compositions create a degeneracy in age determination (see Murphy et
al., 2004). Namely, the small separations $\delta\nu$ change with
the initial composition and age. In the next section, we will
discuss this problem and propose a quantity which may be correlated
with stellar age.

\subsubsection{A new quantity be proposed}
 At the present time, we can know the stellar internal structure and understand the stellar
evolution from oscillation frequencies. Thus asteroseismology
provides a window to ``see" the interior of star. But the
observation of solar-like oscillation is very difficult because of
their small amplitude. So far, we only obtain the knowledge of the
stellar interior from the limited modes ($l=0, 1, 2, 3$) which can
be observed. Many authors proposed some quantity as diagnostic
purposes to probe the stellar internal and constraint the model
parameters (Christensen-Dalsgaard, 1984, 1988, 1993; Ulrich, 1986,
1988; Gough, 1987, 1990a, 2003). For $p$-modes of solar-like stars,
the usual frequency separations are the large separation defined by
equation (2) and the small separation defined by equation (3).
Additionally, Roxburgh (1993) and Roxburgh and Vorontsov (2003)
considered the following separations:
\begin {equation}
\delta_{01}(n)=\frac{1}{8}(\nu_{n-1,0}-4\nu_{n-1,1}+6\nu_{n,0}-4\nu_{n,1}+\nu_{n+1,0})
,
\end {equation}

\begin {equation}
\delta_{10}(n)=-\frac{1}{8}(\nu_{n-1,1}-4\nu_{n-1,0}+6\nu_{n,1}-4\nu_{n+1,0}+\nu_{n+1,1})
,
\end {equation}
and defined the ratios $d_{ij}$ of small to large separations as
follows:
\begin {equation}
\begin{split}
d_{02}(n)=\frac{\delta\nu_{02}(n)}{\Delta\nu_{1}(n)},\hspace{0.5cm}
d_{13}(n)=\frac{\delta\nu_{13}(n)}{\Delta\nu_{0}(n+1)},\\
d_{01}(n)=\frac{\delta\nu_{01}(n)}{\Delta\nu_{1}(n)},\hspace{0.5cm}
d_{10}(n)=\frac{\delta\nu_{10}(n)}{\Delta\nu_{0}(n+1)}.
 \end{split}
\end {equation}
The ratios $d_{ij}$ of small to large separations are independent of
the structure of the outer layers of a star, and therefore provide a
diagnostic of the stellar interior alone.

In addition, Gough (1990a), Monteiro and Thompson (1998), Vauclair
and Th\'{e}ado (2004),  Houdek and Gough (2007a) gave the second
differences $\Delta_{2}\nu_{l}(n)$:

\begin {equation}
\Delta_{2}\nu_{l}(n)=\nu_{n+1,l}+\nu_{n-1,l}-2\nu_{n,l}.
\end {equation}
The second differences $\Delta_{2}\nu_{l}(n)$ can be used to reveal
the variation of the first adiabatic exponent $\gamma_1$ dependent
of the influence of the ionization of helium on the low-degree
acoustic oscillation frequencies in model of solar-type stars.
Recently, Houdek and Gough (2007b) stated that the second
differences can provide a measure of helium abundance and hence
precisely lift the degeneracy between composition and age.

Summarizing the above character separation, we find that
investigation of lifting the degeneracy between the chemical
compositions and the age is interesting. We begin with our
investigation from a well-known asymptotic formula.

The asymptotic formula for the frequency $\nu_{n,l}$ of a stellar
$p$-mode of order $n$ and degree $l$ was given by Tassoul (1980):
\begin {equation}
\nu_{n,l}\simeq(n+\frac{l}{2}+\epsilon)\nu_{0}-[Al(l+1)-B]\nu_{0}^{2}\nu_{n,l}^{-1},
\end {equation}
where, the characteristic $\nu_{0}$ is related to the run of sound
travel time across the stellar diameter; $A$ is a measure of the
sound-speed gradient and most sensitive to conditions in the stellar
core (see Gough and Novotny, 1990b; Gough, 2003;
Christensen-Dalsgaard, 1993; Guenther and Brown, 2004), $\epsilon$
and $B$ are constants which are the functions of the equilibrium
model. It should be noted that the classical asymptotic theory of
Tassoul (1980), although providing good results at the first order
in frequency, does not represent with accuracy the $p$-mode spectrum
of the stars considered. Several authors (e.g., Gabriel, 1989;
Audard and Provost, 1994; Roxburgh and Vorontsov, 2000a, 2000b,
2001) have been discussed the difficulties of the asymptotic theory,
particularly for evolved models with rapid variation in the sound
speed in the core.

Using the equation (2) and the asymptotic formula (10), the large
separation can be written as follows (Gough and Novotny, 1990b):
\begin {equation}
\begin{split}
\Delta\nu_{n,l}&=\nu_{n,l}-\nu_{n-1,l}\\
               &=\nu_{0}-[Al(l+1)-B]\nu_{0}^{2}\frac{\nu_{n-1,l}-\nu_{n,l}}{\nu_{n,l}\cdot\nu_{n-1,l}},
 \end{split}
\end {equation}

Taking the first order of $\nu_{n,l}$ for the $n\gg l$
approximately, we can obtain the result like Gough and Novotny
(1990b) and equation (11) becomes:,

\begin {equation}
\begin{split}
\Delta\nu_{n,l}&\approx\nu_{0}[1-\frac{Al(l+1)-B}{(n+\frac{l}{2}+\epsilon)(n-1+\frac{l}{2}+\epsilon)}]^{-1}\\
               &\approx\nu_{0}.
 \end{split}
\end {equation}
Using the same approximate method, we can obtain the expression of
small separation:

\begin {equation}
\delta\nu_{n,l}=\delta\nu_{l,l+2}\approx\frac{(4l+6)\nu_{0}A}{n+\frac{l}{2}+\epsilon}.
\end {equation}

Due to the small separations are rather sensitive to composition and
therefore to the structure of the core, especially the extreme
sensitivity of the stellar core density stratification to several
parameters (Guenther and Demarque, 2000; Morel et al., 2000), we
define another quantity about the ratio of average small separation
adjacent in $l$:

\begin {equation}
r_{01}=\frac{\langle\delta\nu_{0,2}\rangle}{\langle\delta\nu_{1,3}\rangle}.
\end {equation}

Using the equation (14), we calculate the values of $r_{01}$ and
list it in table 3. Based on the results of numerical calculations,
we plot the ratio $r_{01}$ versus age in Fig. 6. Fortunately, we
find that the ratio $r_{01}$ is tightly correlated with age and
decreases monotonously with age. We think that the likely reason
comes from the perturbation to the gravitational potential,
neglected in the asymptotic relation (10), which affects modes of
the lowest degrees most strongly and which probably increases with
evolution due to the increasing central density. These effects are
most important for modes of the lowest degrees which penetrate most
deeply and hence affect $\delta \nu_{0,2}$ more than $\delta
\nu_{1,3}$, leading to the dependence of $r_{01}$ on age.

From the Fig. 6, the values of $r_{0,1}$ in table 3 and the above
discussion, we can conclude that this quantity $r_{01}$ can lift the
degeneracy between the chemical compositions and age. The analysis
was inspired by Fernandes and Monteiro (2003) and Murphy et al.
(2004). So, we can obtain the $r_{01}$ which may indicate stellar
age, if we consider a frequency ratio. As illustrated in Fig. 6, the
quantity $r_{01}$ is tightly correlate with stellar age over a
substantial range of the remaining parameters, including
composition. At the same time, we need to point out that the range
of variation of this quantity is relatively modest, compared to
likely observational errors. Also, it is clear that the present data
for 70 Ophiuchi A are not adequate to evaluate this quantity. We
think that using this quantity to evaluate the stellar age will be
convenient based on asteroseismic data which will be provided in the
future.
\section{Discussion and conclusions}

The models of 70 Ophiuchi A are obtained by fitting effective
temperature, luminosity, surface metallicity and asteroseismic
observations.

We list a series of possible models in Table 3. So far, we think
that Model 60, Model 125 and Model 126 are the best models. With the
advance in observation, the precise asteroseismic data will provide
more strict constraints on theoretical models. The conclusions of
the paper are:

1. Using the latest asteroseismic observation, we try our best to
construct the best model of 70 Ophiuchi A. So far, we only select
the Model 60, Model 125 and Model 126, which can be fit for the
observation, as the optimum models. These models are correspond to
the radius of this star is about 0.860 -- 0.865 $R_{\odot}$
   and the age is about 6.8 -- 7.0 Gyr with mass 0.89 -- 0.90
   $M_{\odot}$.

2. By calculating many theoretical models, we want to use the
theoretical frequencies to compare with observational frequencies
and help to definitely validate the observational data.

3. We obtain a new quantity $r_{01}$ which can lift the degeneracy
between the initial compositions and the stellar age. By
calculation, we prove that it can be valuable for the indication of
the stellar age.

4. Important point is that we test our stellar structure and
evolution theory. Meanwhile we find the confrontation of
observations and theoretical models. Actually we have neglect some
important effects, such as rotation and magnetic field, which can
impact on the internal structure and evolution. Thus, the theory
models which we have constructed can not fit observation perfectly.
The detailed comparisons of individual mode frequencies will also
require taking into account the effects of turbulence in the outer
convective unstable layers in the stellar models, which shift the
observed frequencies. The parameterization of turbulence tested on
the Sun by Li et al. (2002) can be applied to models for solar-like
stars as well. This parameterization can be extended to other stars
by using the three-dimensional radiative hydrodynamic simulations of
Robinson et al. (2003), which are based on the same microscopic
physics and can readily be parameterized in the YREC stellar
evolution code.

\vspace{2mm}
{\bf Acknowledgements} \vspace{1mm}

We are grateful to anonymous referees for their constructive
suggestions and valuable remarks to improve the manuscript. This
work was supported by The Ministry of Science and Technology of the
People's republic of China through grant 2007CB815406, and by NSFC
grants 10173021, 10433030, 10773003, and 10778601.

\vspace{2mm}
{\bf References} \vspace{1mm}

 Alexander, D. R., Ferguson, J. W., 1994. ApJ 437, 879.

 Audard, N., Provost, J., 1994. A\&A 282, 73.

 Bahcall, J. N., Pinsonneault, M. H., Wasserburg, G. J., 1995. RvMP 67, 781.

 Batten, A. H., Fletcher, J. M., Campbell, B., 1984. PASP 96, 903.

 Bedding, T. R., Butler, R. P., Kjeldsen, H., Baldry, I. K.,
  O'Toole, S. J., Tinney, C. G., Marcy, G. W., Kienzle, F.,
  Carrier, F., 2001. ApJ 549, L105.

 Bedding, T. R., Kjeldsen, H.; Butler, R. P., McCarthy, C., Marcy, G.
 W., O'Toole, S. J., Tinney, C. G., Wright, J. T., 2004. ApJ 614, 380.

 Bouchy, F., Carrier, F., 2002. A\&A 390, 205.

 Bouchy, F., Bazot, M., Santos, N. C., Vauclair, S., Sosnowska, D., 2005. A\&A 440,
 609.

 Carrier, F., Bouchy, F., Kienzle, F., Bedding, T. R., Kjeldsen, H., Butler, R. P., Baldry, I.
 K., O'Toole, S. J., Tinney, C. G., Marcy, G. W., 2001. A\&A 378, 142.

 Carrier, F., Bourban, G., 2003a. A\&A 406, L23.

 Carrier, F., Bouchy, F., Eggenberger, P., 2003b. In Thompson, M. J., Cunha, M. S.,
   \& Monteiro, M. J. P. F. G., editors, Asteroseismology Across the HR Diagram, page P311. Kluwer.

 Carrier, F., Eggenberger, P., Bouchy, F., 2005a. A\&A 434, 1085.

 Carrier, F., Eggenberger, P., D¡¯Alessandro, A., Weber, L., 2005b. NewA 10,
 315.

 Carrier, F., Eggenberger, P., 2006. A\&A 450, 695.

 Christensen-Dalsgaard, J., 1984. in Mangeney A., Praderie, F., eds, Space Research Prospects in Stellar
   Activity and Variability, Paris Observatory Press, Paris, P11.

 Christensen-Dalsgaard, J., 1988. in Advances in Helio and Asteroseismology, ed. J. Christensen-Dalsgaard,
    \& S. Frandsen(Reidel), 295.

 Christensen-Dalsgaard, J., 1993. in Seismic Investigation of the Sun
    and Stars, ed. T. M. Bron, A.S.P. Conf. Ser 42, 347.

 Eggenberger, P., Carrier, F., Bouchy, F., Blecha, A., 2004a. A\&A 422,
 247.

 Eggenberger, P., Charbonnel, C., Talon, S., Meynet, G., Maeder,
     A., Carrier, F., Bourban, G., 2004b. A\&A 417, 235.

 Eggenberger, P., Carrier, F., Bouchy, F., 2005. NewA 10, 195.

 Fernandes, J., Lebreton, Y., Baglin, A., Morel, P., 1998. A\&A 338,
 455.

 Fernandes, J., Monteiro, M. J. P. G., 2003. A\&A 399, 243.

 Gabriel, M., 1989. A\&A 226, 278.

 Gough, D. O., 1987. Nature 326, 257.

 Gough, D. O., 1990a. in Progress of Seismology of the Sun and Stars, Proc. Oji International Seminar Hakone
    (Japan: Springer Verlag), Lect. Notes Phys., 367, 283.

 Gough, D. O., Novotny, E., 1990b. SoPh 128, 143.

 Gough, D. O., 2003. Ap\&SS 284, 165.

 Grevesse. N., Sauval, A. J., 1998. SSRv 85, 161.

 Gray, D. H., Johanson, H., 1991. PASP 103, 439.

 Guenther, D. B., Demarque, P., Kim, Y.-C., Pinsonneault, M. H., 1992. ApJ 387,
 372.

 Guenther, D. B., 1994. ApJ 422, 400.

 Guenther, D. B., 1998. in Proc. SOHO 6/GONG 98 Workshop, Structure and Dynamics
of the Interior of the Sun and Sun-like Stars, ed S. Korzennik \& A.
Wilson (ESA SP-418), 375.

 Guenther, D. B., Demarque, P., 2000. ApJ 531, 503.

 Guenther, D. B., Brown, K. I. T., 2004. ApJ 600, 419.

 Heintz, W. D., 1988. JRASC 82, 140.

 Houdek, G., Gough, D. O., 2007a. MNRAS 375, 861.

 Houdek, G., Gough, D. O., 2007b. AIPC 984, 219.

 Iglesias, C. A., Rogers, F. J., 1996. ApJ 464, 943.

 Kervella, P., Th\'{e}venin, F., Morel, P., Bord\'{e}, P., Di Folco, E., 2003a. A\&A 408,
 681.

 Kervella, P., Th\'{e}venin, F., S\'{a}gransan, D., Berthomieu, G., Lopez, B., Morel, P., Provost, J., 2003b. A\&A 404,
 1087.

 Kervella, P., Th\'{e}venin, F., Morel, P., Berthomieu, G., Bord\'{e}, P., Provost, J., 2004. A\&A 413,
 251.

Kjeldsen, H., Bedding, T. R., Baldry, I. K., Bruntt, H., Butler, R.
P., Fischer, D. A., Frandsen, S., Gates, E. L., Grundahl, F., Lang,
K., Marcy, G. W., Misch, A., Vogt, S. S., 2003. AJ 126, 1483.

 Kjeldsen, H., Bedding, T. R., Butler, R. P., Christensen-Dalsgaard, J., Kiss, L.
 L., McCarthy, C., Marcy, G. W., Tinney, C. G., Wright, J. T., 2005. ApJ 635,
 1281.

 Li, L. H., Robinson, F. J., Demarque, P., Sofia, S., Guenther, D. B., 2002. ApJ 567,
 1192.

 Marti\'{c}, M., Lebrun, J.-C., Appourchaux, T., Korzennik, S. G., 2004a. A\&A 418,
 295.

 Marti\'{c}, M., Lebrun, J. C., Appourchaux, T., Schmitt, J., 2004b. In Danesy, D., editor, SOHO 14/GONG 2004 Workshop,
    Helio- and Asteroseismology: Towards a Golden Future, ESA SP-559, page
    563.

 Miglio, A., Montalb\'{a}n, J., 2005. A\&A 441, 615.

 Monteiro, M. J. P. F. G., Thompson, M. J., 1998. in New Eyes to See Inside the Sun and Stars (Dordrecht: Kluwer), ed. F. L. Deubner, J.
   Christensen-Dalsgaard, \& D. W. Kurtz, Proc. IAU Symp., 185, 317.

 Morel, P., 1997. A\&AS 124, 597.

 Morel, P., Baglin, A., 1999. A\&A 345, 156.

 Morel, P., Provost, J., Lebreton, Y., Th\'{e}venin, F., Berthomieu, G., 2000. A\&A 363,
 675.

Mosser, B., Bouchy, F., Catala, C., Michel, E., Samadi, R.,
Th\'{e}venin, F., Eggenberger, P., Sosnowska, D., Moutou, C.,
Baglin, A., 2005. A\&A 431, L13.

 Murphy, E. J., Demarque, P., Guenther, D. B., 2004. ApJ 605, 472.

 Perrin, M. N., Cayrel, de S. G., Cayrel, R., 1975. A\&A 39, 97.

 Peterson, R., 1978. ApJ 224, 595.

 Pourbaix, D., 2000. A\&AS 145, 215.

 Provost, J., Martic, M., Berthomieu, G., 2004. ESA SP. 559, 594.

Provost, J., Berthomieu, G., Bigot, L., Morel, P., 2005. A\&A 432,
 225.

 Provost, J., Berthomieu, G., Marti\'{c}, M., Morel, P., 2006. A\&A 460,
 759.

 Robinson, F. J., Demarque, P., Li, L. H., Sofia, S., Kim, Y.-C.,
  Chan, K. L., Guenther, D. B., 2003. MNRAS 340, 923.

 Rogers, F. J., Nayfonov, A., 2002. ApJ 576, 1064.

 Roxburgh, I. W., 1993. in PRISMA, Report of Phase A Study, ed. T.
   Approurchaux, et al., ESA 93, 31.

  Roxburgh, I. W., Vorontsov, S. V., 2000a. MNRAS 317, 141.

Roxburgh, I. W., Vorontsov, S. V., 2000b. MNRAS 317, 151.

Roxburgh, I. W., Vorontsov, S. V., 2001. MNRAS 322, 85.

 Roxburgh, I. W., Vorontsov, S. V., 2003. A\&A 411, 215.

 Tassoul, M., 1980. ApJS 43, 469.

 Thoul, A. A., Bahcall, J. N., Loeb, A., 1994. ApJ 421, 828.

 Th\'{e}venin, F., Provost, J., Morel, P., Berthomieu, G., Bouchy, F., Carrier,
 F., 2002. A\&A 392, 9.

 Ulrich, R. K., 1986. ApJ 306, L37.

 Ulrich, R. K., 1988. IAUS 123, 299.

 Vauclair, S., Th\'{e}ado, S., 2004. A\&A 425, 179.


\center

 {\bf \Huge Online Material}
  \begin{table*}
\caption{Model parameters (M: mass; $Y_{i}$: initial Helium
abundance; $Z_{i}$: initial heavy element abundance; $\log T_{eff}$:
log-effective temperature; $\log L$: log-luminosity; R: radius;
$\langle\Delta\nu_{l}\rangle$:
$\frac{1}{21}\sum^{30}_{n=10}\Delta\nu_{n,l}$;
$\langle\Delta\nu\rangle$:
$\frac{1}{4}\Sigma_{l=0}^{3}\langle\Delta\nu_{l}\rangle$;
$\langle\delta\nu_{02}\rangle$: $\frac{1}{21}\sum^{30}_{n=10}
\delta\nu_{n,0}$; $\langle\delta\nu_{13}\rangle$:
$\frac{1}{21}\sum^{30}_{n=10}\delta\nu_{n,1}$; $r_{01}:
\frac{\langle\delta\nu_{02}\rangle}{\langle\delta\nu_{13}\rangle}$).}
 \tabcolsep0.08in
\tiny{ \[
\begin{tabular}{c c c c  c c c c c c c c c c c c c}
\hline\hline
     & M            &    &                    &             &$\log L$        &R        &Age  &       &              &                    &                   &&& & &\\
Model &($M_{\odot}$)& $Y_{i}$&$Z_{i}$&$\log T_{eff}$&($L_{\odot}$)&($R_{\odot}$) &(Gyr)&$\langle\Delta\nu_{0}\rangle$&$\langle\Delta\nu_{1}\rangle$ &$\langle\Delta\nu_{2}\rangle$&$\langle\Delta\nu_{3}\rangle$&$\langle\Delta\nu\rangle$& $\langle\delta\nu_{02}\rangle$& $\langle\delta\nu_{13}\rangle$&$\langle\Delta\nu_{r}\rangle$&$r_{01}$ \\
\hline

1......&0.85&0.27&0.017&3.7263&-0.319&0.8155&5.15&170.84&170.91&171.06&171.40&171.05&12.46&21.23&125.9677&0.5869\\
2......&0.85&0.27&0.017&3.7270&-0.310&0.8215&5.55&168.81&169.07&169.22&169.56&169.16&12.04&20.65&125.9532&0.5831\\
3......&0.85&0.27&0.017&3.7276&-0.301&0.8277&5.95&167.11&167.19&167.34&167.68&167.33&11.61&20.06&126.0037&0.5788\\
4......&0.85&0.28&0.019&3.7271&-0.319&0.8125&5.05&171.86&171.93&172.07&172.41&172.06&12.48&21.28&126.0129&0.5865\\
5......&0.85&0.28&0.019&3.7276&-0.312&0.8170&5.35&170.46&170.53&170.67&171.01&170.66&12.17&20.84&126.0274&0.5840\\
6......&0.86&0.27&0.017&3.7277&-0.319&0.8104&3.90&173.46&173.50&173.65&173.99&173.65&13.50&22.65&126.6846&0.5960\\
7......&0.86&0.27&0.018&3.7265&-0.319&0.8150&4.60&172.05&172.11&172.26&172.60&172.25&12.87&21.79&126.7347&0.5906\\
8......&0.86&0.27&0.018&3.7271&-0.311&0.8201&4.95&170.46&170.52&170.67&171.01&170.66&12.51&21.28&126.7453&0.5879\\
9......&0.86&0.27&0.018&3.7276&-0.304&0.8246&5.25&169.07&169.14&169.28&169.62&169.27&12.19&20.84&126.7491&0.5849\\
10.....&0.86&0.28&0.020&3.7273&-0.319&0.8119&4.50&173.11&173.16&173.30&173.63&173.30&12.91&21.84&126.7805&0.5911\\
11.....&0.86&0.28&0.020&3.7276&-0.315&0.8140&4.65&172.43&172.49&172.63&172.96&172.62&12.75&21.63&126.7733&0.5895\\
12.....&0.87&0.26&0.017&3.7258&-0.319&0.8170&4.10&172.40&172.44&172.59&172.94&172.59&13.30&22.36&127.4526&0.5948\\
13.....&0.87&0.26&0.017&3.7267&-0.307&0.8248&4.65&169.96&170.02&170.17&170.51&170.16&12.74&21.57&127.4619&0.5906\\
14.....&0.87&0.26&0.017&3.7276&-0.295&0.8330&5.20&167.46&167.52&167.67&168.01&167.66&12.17&20.78&127.4668&0.5857\\
15.....&0.87&0.26&0.018&3.7246&-0.319&0.8213&4.80&171.09&171.15&171.29&171.64&171.29&12.69&21.52&127.4925&0.5897\\
16.....&0.87&0.26&0.018&3.7261&-0.300&0.8343&5.70&167.10&167.17&167.32&167.66&167.31&11.78&20.25&127.4986&0.5817\\
17.....&0.87&0.26&0.018&3.7276&-0.279&0.8487&6.60&162.90&163.00&163.13&163.48&163.12&10.84&18.94&127.5377&0.5723\\
18.....&0.87&0.27&0.019&3.7266&-0.319&0.8143&4.05&173.30&173.34&173.49&173.83&173.49&13.30&22.35&127.4827&0.5951\\
19.....&0.87&0.27&0.019&3.7276&-0.306&0.8230&4.65&170.59&170.64&170.79&171.12&170.78&12.67&21.49&127.5078&0.5896\\
20.....&0.87&0.27&0.020&3.7255&-0.319&0.8187&4.75&171.96&172.01&172.15&172.49&172.15&12.68&21.51&127.5247&0.5895\\
21.....&0.87&0.27&0.020&3.7266&-0.304&0.8282&5.40&169.02&169.09&169.22&169.56&169.22&12.01&20.58&127.5424&0.5836\\
22.....&0.87&0.27&0.020&3.7276&-0.290&0.8375&6.00&166.21&166.29&166.42&166.75&166.41&11.38&19.70&127.5430&0.5777\\
23.....&0.87&0.28&0.021&3.7275&-0.319&0.8110&3.95&174.43&174.46&174.61&174.94&174.61&13.34&22.42&127.5265&0.5950\\
24.....&0.87&0.28&0.021&3.7276&-0.318&0.8117&4.00&174.20&174.24&174.38&174.71&174.38&13.29&22.35&127.5234&0.5946\\
25.....&0.88&0.26&0.017&3.7271&-0.319&0.8124&2.90&174.85&174.88&175.04&175.37&175.03&14.30&23.72&128.1644&0.6029\\
26.....&0.88&0.26&0.017&3.7277&-0.311&0.8172&3.25&173.31&173.34&173.49&173.83&173.49&13.94&23.22&128.1643&0.6003\\
27.....&0.88&0.26&0.018&3.7260&-0.319&0.8170&3.60&173.43&173.46&173.62&173.95&173.61&13.67&22.85&128.2058&0.5982\\
28.....&0.88&0.26&0.018&3.7268&-0.308&0.8239&4.10&171.23&171.28&171.43&171.76&171.42&13.16&22.14&128.1956&0.5944\\
29.....&0.88&0.26&0.018&3.7276&-0.297&0.8312&4.60&168.98&169.03&169.18&169.52&169.17&12.64&21.42&128.1981&0.5901\\
30.....&0.88&0.26&0.019&3.7247&-0.320&0.8206&4.25&172.32&172.36&172.51&172.84&172.50&13.11&22.07&128.2290&0.5940\\
31.....&0.88&0.26&0.019&3.7256&-0.309&0.8276&4.75&170.14&170.20&170.34&170.68&170.34&12.61&21.38&128.2470&0.5898\\
32.....&0.88&0.26&0.019&3.7264&-0.298&0.8349&5.25&167.90&167.97&168.11&168.44&168.10&12.10&20.66&128.2388&0.5857\\
33.....&0.88&0.26&0.019&3.7272&-0.287&0.8427&5.75&165.61&165.69&165.83&166.16&165.82&11.58&19.94&128.2763&0.5807\\
34.....&0.88&0.26&0.019&3.7277&-0.281&0.8467&6.00&164.45&164.53&164.67&165.00&164.66&11.31&19.57&128.2869&0.5779\\
35.....&0.88&0.27&0.020&3.7267&-0.319&0.8135&3.50&174.60&174.63&174.78&175.11&174.78&13.72&22.93&128.2413&0.5983\\
36.....&0.88&0.27&0.020&3.7272&-0.314&0.8170&3.75&173.49&173.52&173.67&174.00&173.67&13.47&22.57&128.2501&0.5968\\
37.....&0.88&0.27&0.020&3.7277&-0.307&0.8212&4.05&172.13&172.17&172.32&172.65&172.31&13.16&22.14&128.2283&0.5944\\
38.....&0.88&0.27&0.021&3.7257&-0.319&0.8179&4.20&173.23&173.27&173.42&173.75&173.41&13.11&22.07&128.2698&0.5940\\
39.....&0.88&0.27&0.021&3.7259&-0.316&0.8200&4.35&172.57&172.61&172.76&173.09&172.75&12.95&21.86&128.2741&0.5924\\
40.....&0.88&0.27&0.021&3.7265&-0.308&0.8250&4.70&171.00&171.05&171.19&171.52&171.19&12.59&21.36&128.2801&0.5894\\
41.....&0.88&0.27&0.021&3.7273&-0.297&0.8325&5.20&168.70&168.76&168.90&169.22&168.89&12.07&20.63&128.2863&0.5851\\
42.....&0.88&0.27&0.021&3.7277&-0.292&0.8356&5.40&167.76&167.83&167.97&168.29&167.96&11.86&20.34&128.2932&0.5831\\
43.....&0.89&0.26&0.018&3.7272&-0.319&0.8122&2.40&175.96&175.97&176.13&176.46&176.13&14.68&24.23&128.9222&0.6059\\
44.....&0.89&0.26&0.018&3.7277&-0.313&0.8155&2.65&174.87&174.88&175.04&175.37&175.04&14.43&23.88&128.9060&0.6043\\
45.....&0.89&0.26&0.019&3.7260&-0.319&0.8162&3.05&174.71&174.73&174.89&175.22&174.88&14.09&23.41&128.9541&0.6019\\
46.....&0.89&0.26&0.019&3.7265&-0.313&0.8202&3.35&173.42&173.45&173.60&173.93&173.60&13.79&23.00&128.9524&0.5996\\
47.....&0.89&0.26&0.019&3.7270&-0.306&0.8244&3.65&172.10&172.13&172.28&172.61&172.28&13.49&22.57&128.9561&0.5977\\
48.....&0.89&0.26&0.019&3.7276&-0.299&0.8294&4.00&170.54&170.58&170.73&171.06&170.72&13.13&22.07&128.9527&0.5949\\
49.....&0.89&0.26&0.020&3.7249&-0.319&0.8205&3.75&173.36&173.39&173.54&173.87&173.54&13.48&22.56&128.9785&0.5975\\
50.....&0.89&0.26&0.020&3.7254&-0.313&0.8246&4.05&172.07&172.11&172.26&172.59&172.25&13.18&22.15&128.9805&0.5950\\
51.....&0.89&0.26&0.020&3.7257&-0.308&0.8274&4.25&171.20&171.24&171.39&171.72&171.38&12.98&21.86&128.9833&0.5938\\
52.....&0.89&0.26&0.020&3.7266&-0.298&0.8347&4.75&168.98&169.03&169.17&169.50&169.17&12.47&21.16&129.0087&0.5893\\
53.....&0.89&0.26&0.020&3.7274&-0.286&0.8422&5.25&166.71&166.78&166.91&167.24&166.91&11.95&20.44&129.0046&0.5846\\
54.....&0.89&0.26&0.020&3.7276&-0.283&0.8446&5.40&166.02&166.09&166.22&166.55&166.22&11.19&20.22&129.0209&0.5534\\

\end{tabular}\]}
\end{table*}
\addtocounter{table}{-1}
\begin{table*}
 \caption{-Continued}
 \tabcolsep0.08in
\tiny{ \[
\begin{tabular}{c c c c  c c c c c c c c c c c c c}
\hline\hline
      & M            &    &                    &             &$\log L$        &R        &Age  &       &              &                    &                   &&& & &\\
Model &($M_{\odot}$)& $Y_{i}$&$Z_{i}$&$\log T_{eff}$&($L_{\odot}$)&($R_{\odot}$) &(Gyr)&$\langle\Delta\nu_{0}\rangle$&$\langle\Delta\nu_{1}\rangle$ &$\langle\Delta\nu_{2}\rangle$&$\langle\Delta\nu_{3}\rangle$&$\langle\Delta\nu\rangle$& $\langle\delta\nu_{02}\rangle$& $\langle\delta\nu_{13}\rangle$&$\langle\Delta\nu_{r}\rangle$&$r_{01}$ \\
\hline

55.....&0.89&0.26&0.021&3.7238&-0.319&0.8247&4.45&172.08&172.13&172.27&172.60&172.27&12.88&21.73&129.0190&0.5927\\
56.....&0.89&0.26&0.021&3.7246&-0.309&0.8316&4.95&169.94&169.99&170.13&170.46&170.13&12.38&21.04&129.0187&0.5884\\
57.....&0.89&0.26&0.021&3.7255&-0.298&0.8389&5.45&167.73&167.80&167.94&168.26&167.93&11.88&20.34&129.0309&0.5841\\
58.....&0.89&0.26&0.021&3.7263&-0.286&0.8466&5.95&165.47&165.55&165.67&166.00&165.67&11.36&19.63&129.0510&0.5787\\

59.....&0.89&0.26&0.021&3.7271&-0.275&0.8547&6.45&163.14&163.24&163.36&163.69&163.35&10.84&18.90&129.0743&0.5735\\
$\mathbf{60}$.....&$\mathbf{0.89}$&$\mathbf{0.26}$&$\mathbf{0.021}$&$\mathbf{3.7277}$&
$\mathbf{-0.267}$&$\mathbf{0.8606}$&$\mathbf{6.80}$&$\mathbf{161.48}$&$\mathbf{161.59}$&$\mathbf{161.70}$
&$\mathbf{162.04}$&$\mathbf{161.70}$&$\mathbf{10.48}$&$\mathbf{18.39}$&$\mathbf{129.0958}$&$\mathbf{0.5699}$\\
61.....&0.89&0.27&0.021&3.7269&-0.319&0.8132&3.00&175.72&175.74&175.89&176.22&175.89&14.11&23.44&128.9844&0.6020\\
62.....&0.89&0.27&0.021&3.7276&-0.309&0.8195&3.45&173.71&173.74&173.89&174.21&173.88&13.65&22.80&128.9951&0.5987\\
63.....&0.90&0.26&0.019&3.7273&-0.319&0.8118&1.90&177.12&177.13&177.29&177.62&177.29&15.07&24.75&129.6755&0.6089\\
64.....&0.90&0.26&0.019&3.7276&-0.315&0.8144&2.10&176.24&176.26&176.42&176.74&176.41&14.87&24.47&129.6522&0.6077\\
65.....&0.90&0.26&0.020&3.7261&-0.319&0.8159&2.55&175.82&175.84&175.99&176.32&175.99&14.47&23.92&129.7010&0.6049\\
66.....&0.90&0.26&0.020&3.7270&-0.308&0.8233&3.10&173.44&173.47&173.62&173.94&173.61&13.92&23.15&129.6916&0.6013\\
67.....&0.90&0.26&0.020&3.7276&-0.300&0.8282&3.45&171.90&171.93&172.08&172.40&172.07&13.56&22.66&129.6904&0.5984\\
68.....&0.90&0.26&0.021&3.7250&-0.319&0.8203&3.25&174.43&174.46&174.61&174.93&174.60&13.85&23.05&129.7189&0.6009\\
69.....&0.90&0.26&0.021&3.7254&-0.314&0.8237&3.50&173.37&173.39&173.54&173.87&173.54&13.60&22.71&129.7338&0.5989\\
70.....&0.90&0.26&0.021&3.7259&-0.308&0.8271&3.75&172.29&172.32&172.46&172.79&172.46&13.35&22.36&129.7255&0.5970\\
71.....&0.90&0.26&0.021&3.7267&-0.298&0.8342&4.25&170.09&170.13&170.27&170.60&170.27&12.84&21.66&129.7309&0.5928\\
72.....&0.90&0.26&0.021&3.7276&-0.285&0.8424&4.80&167.62&167.67&167.81&168.13&167.80&12.28&20.87&129.7387&0.5884\\
73.....&0.91&0.26&0.020&3.7274&-0.319&0.8112&1.40&178.34&178.35&178.51&178.84&178.51&15.46&25.28&130.4231&0.6116\\
74.....&0.91&0.26&0.020&3.7277&-0.315&0.8138&1.60&177.46&177.47&177.63&177.95&177.62&15.26&25.00&130.3972&0.6104\\
75.....&0.91&0.26&0.021&3.7262&-0.319&0.8154&2.05&176.98&176.99&177.15&177.47&177.14&14.85&24.43&130.4286&0.6079\\
76.....&0.91&0.26&0.021&3.7270&-0.309&0.8221&2.55&174.82&174.84&174.99&175.31&174.99&14.36&23.74&130.4368&0.6049\\
77.....&0.91&0.26&0.021&3.7277&-0.300&0.8277&2.95&173.07&173.09&173.24&173.56&173.24&13.95&23.17&130.4541&0.6021\\
78.....&0.85&0.29&0.022&3.7244&-0.353&0.7914&4.05&178.88&178.91&179.06&179.39&179.06&13.56&22.78&126.0645&0.5953\\
79.....&0.85&0.29&0.022&3.7261&-0.332&0.8045&5.00&174.56&174.62&174.75&175.08&174.75&12.59&21.42&126.0974&0.5878\\
80.....&0.85&0.29&0.022&3.7277&-0.310&0.8195&6.00&169.78&169.87&169.99&170.32&169.99&11.52&19.94&126.1092&0.5777\\
81.....&0.86&0.28&0.022&3.7248&-0.323&0.8176&5.70&171.40&171.47&171.60&171.93&171.60&11.91&20.46&126.8611&0.5821\\
82.....&0.86&0.28&0.022&3.7262&-0.305&0.8297&6.50&167.68&167.77&167.89&168.22&167.89&11.08&19.31&126.8839&0.5738\\
83.....&0.86&0.28&0.022&3.7277&-0.284&0.8446&7.40&163.28&163.40&163.51&163.85&163.51&10.13&17.97&126.9173&0.5637\\
84.....&0.87&0.28&0.022&3.7263&-0.321&0.8141&4.55&173.48&173.53&173.67&174.00&173.67&12.82&21.70&127.5679&0.5908\\
85.....&0.87&0.28&0.022&3.7270&-0.312&0.8199&4.95&171.66&171.72&171.85&172.18&171.85&12.41&21.13&127.5824&0.5873\\
86.....&0.87&0.28&0.022&3.7277&-0.303&0.8260&5.35&169.78&169.84&169.98&170.30&169.97&11.99&20.54&127.5975&0.5837\\
87.....&0.88&0.27&0.022&3.7244&-0.322&0.8208&4.80&172.37&172.43&172.57&172.89&172.56&12.61&21.38&128.3205&0.5898\\
88.....&0.88&0.27&0.022&3.7261&-0.299&0.8354&5.80&167.89&167.97&168.09&168.42&168.09&11.59&19.96&128.3464&0.5807\\
89.....&0.88&0.27&0.022&3.7278&-0.276&0.8516&6.80&162.92&163.02&163.13&163.47&163.13&10.49&18.42&128.1998&0.5695\\
90.....&0.89&0.26&0.022&3.7244&-0.297&0.8439&6.20&166.32&166.41&166.53&166.86&166.53&11.26&19.48&129.1008&0.5780\\
91.....&0.89&0.26&0.022&3.7251&-0.288&0.8500&6.60&164.54&164.64&164.76&165.08&164.75&10.86&18.92&129.1082&0.5740\\
92.....&0.89&0.26&0.022&3.7258&-0.279&0.8564&7.00&162.70&162.81&162.92&163.25&162.92&10.45&18.35&129.1188&0.5695\\
93.....&0.89&0.27&0.022&3.7257&-0.320&0.8171&3.65&174.52&174.55&174.70&175.02&174.69&13.53&22.63&129.0271&0.5979\\
94.....&0.89&0.27&0.022&3.7267&-0.308&0.8248&4.20&172.10&172.14&172.28&172.60&172.28&12.97&21.86&129.0500&0.5933\\
95.....&0.89&0.27&0.022&3.7277&-0.295&0.8336&4.80&169.36&169.42&169.55&169.87&169.55&12.35&20.99&129.0430&0.5884\\
96.....&0.90&0.26&0.022&3.7245&-0.312&0.8294&4.30&171.64&171.68&171.82&172.14&171.82&12.90&21.73&129.7836&0.5936\\
97.....&0.90&0.26&0.022&3.7262&-0.291&0.8431&5.25&167.49&167.55&167.68&168.00&167.68&11.94&20.41&129.8075&0.5850\\
98.....&0.90&0.26&0.022&3.7277&-0.269&0.8581&6.20&163.12&163.21&163.33&163.65&163.33&10.96&19.04&129.8293&0.5756\\
99.....&0.90&0.27&0.022&3.7274&-0.314&0.8164&2.75&175.74&175.76&175.91&176.23&175.91&14.23&23.59&129.7612&0.6032\\
100....&0.90&0.27&0.022&3.7276&-0.312&0.8177&2.85&175.32&175.34&175.49&175.81&175.49&14.14&23.45&129.7608&0.6030\\
101....&0.90&0.27&0.022&3.7277&-0.311&0.8183&2.90&175.10&175.12&175.27&175.59&175.27&14.09&23.39&129.7408&0.6023\\
102....&0.91&0.26&0.022&3.7251&-0.320&0.8194&2.70&175.76&175.78&175.93&176.25&175.93&14.27&23.62&130.4920&0.6041\\
103....&0.91&0.26&0.022&3.7264&-0.304&0.8302&3.50&172.34&172.36&172.51&172.83&172.51&13.48&22.51&130.4933&0.5988\\
104....&0.91&0.26&0.022&3.7277&-0.286&0.8418&4.30&168.79&168.83&168.97&169.29&168.97&12.66&21.38&130.5037&0.5921\\
105....&0.92&0.26&0.022&3.7271&-0.309&0.8222&2.10&175.80&175.81&175.97&176.28&175.97&14.69&24.18&131.1912&0.6075\\
106....&0.92&0.26&0.022&3.7274&-0.306&0.8243&2.25&175.16&175.17&175.32&175.64&175.32&14.54&23.97&131.2077&0.6066\\
107....&0.92&0.26&0.022&3.7277&-0.303&0.8263&2.40&174.51&174.52&174.67&174.99&174.67&14.39&23.76&131.1973&0.6056\\
108....&0.85&0.29&0.023&3.7258&-0.320&0.8169&6.25&170.64&170.72&170.84&171.17&170.84&11.41&19.79&126.1371&0.5766\\
109....&0.85&0.29&0.023&3.7265&-0.312&0.8223&6.60&168.97&169.06&169.17&169.51&169.18&11.04&19.28&126.1521&0.5726\\
110....&0.85&0.29&0.023&3.7271&-0.302&0.8288&7.00&167.00&167.11&167.21&167.55&167.22&10.61&18.68&126.1720&0.5680\\
111....&0.86&0.28&0.023&3.7244&-0.314&0.8276&6.80&168.35&168.44&168.55&168.89&168.56&10.93&19.10&126.9069&0.5723\\
112....&0.86&0.28&0.023&3.7246&-0.312&0.8291&6.90&167.88&167.98&168.09&168.42&168.09&10.83&18.96&126.8973&0.5712\\

\end{tabular}\]}
\end{table*}
\addtocounter{table}{-1}
\begin{table*}
 \caption{-Continued}
 \tabcolsep0.08in
\tiny{ \[
\begin{tabular}{c c c c  c c c c c c c c c c c c c}
\hline\hline
      & M            &    &                    &             &$\log L$        &R        &Age  &       &              &                    &                   &&& & &\\
Model &($M_{\odot}$)& $Y_{i}$&$Z_{i}$&$\log T_{eff}$&($L_{\odot}$)&($R_{\odot}$) &(Gyr)&$\langle\Delta\nu_{0}\rangle$&$\langle\Delta\nu_{1}\rangle$ &$\langle\Delta\nu_{2}\rangle$&$\langle\Delta\nu_{3}\rangle$&$\langle\Delta\nu\rangle$& $\langle\delta\nu_{02}\rangle$& $\langle\delta\nu_{13}\rangle$&$\langle\Delta\nu_{r}\rangle$&$r_{01}$ \\
\hline

113....&0.86&0.28&0.023&3.7248&-0.310&0.8307&7.00&167.42&167.52&167.63&167.96&167.63&10.72&18.82&126.9165&0.5696\\
114....&0.87&0.28&0.023&3.7258&-0.314&0.8228&5.55&170.78&170.85&170.98&171.30&170.98&11.91&20.44&127.6106&0.5827\\

115....&0.87&0.28&0.023&3.7268&-0.300&0.8319&6.15&168.00&168.08&168.20&168.52&168.20&11.29&19.57&127.6241&0.5769\\
116....&0.87&0.28&0.023&3.7278&-0.286&0.8417&6.75&165.10&165.20&165.31&165.64&165.31&10.65&18.68&127.6542&0.5701\\
117....&0.88&0.27&0.023&3.7246&-0.305&0.8358&6.25&167.81&167.89&168.01&168.34&168.01&11.27&19.52&128.3774&0.5774\\
118....&0.88&0.27&0.023&3.7252&-0.297&0.8411&6.60&166.24&166.33&166.44&166.77&166.45&10.91&19.03&128.3971&0.5733\\
119....&0.88&0.27&0.023&3.7258&-0.288&0.8475&7.00&164.37&164.47&164.58&164.91&164.59&10.50&18.46&128.4142&0.5688\\
120....&0.89&0.27&0.023&3.7257&-0.307&0.8299&4.95&170.55&170.60&170.74&171.06&170.74&12.32&20.96&129.0845&0.5878\\
121....&0.89&0.27&0.023&3.7267&-0.293&0.8388&5.55&167.83&167.90&168.03&168.35&168.03&11.70&20.10&129.0846&0.5821\\
122....&0.89&0.27&0.023&3.7277&-0.280&0.8484&6.15&165.02&165.11&165.22&165.54&165.22&11.07&19.22&129.1111&0.5760\\
123....&0.90&0.26&0.023&3.7246&-0.297&0.8436&5.70&167.37&167.44&167.57&167.89&167.57&11.62&19.96&129.8378&0.5821\\
124....&0.90&0.26&0.023&3.7257&-0.282&0.8535&6.35&164.49&164.57&164.69&165.01&164.69&10.96&19.05&129.8591&0.5753\\
$\mathbf{125}$....&$\mathbf{0.90}$&$\mathbf{0.26}$&$\mathbf{0.023}$&$\mathbf{3.7267}$&$\mathbf{-0.269}$&$\mathbf{0.8633}$&
$\mathbf{6.95}$&$\mathbf{161.71}$&$\mathbf{161.81}$&$\mathbf{161.92}$&$\mathbf{162.25}$&$\mathbf{161.92}$&$\mathbf{10.34}$
&$\mathbf{18.18}$&$\mathbf{129.8802}$&$\mathbf{0.5688}$\\
$\mathbf{126}$....&$\mathbf{0.90}$&$\mathbf{0.26}$&$\mathbf{0.023}$&$\mathbf{3.7268}$&$\mathbf{-0.267}$&$\mathbf{0.8641}$
&$\mathbf{7.00}$&$\mathbf{161.47}$&$\mathbf{161.57}$&$\mathbf{161.68}$&$\mathbf{162.01}$&$\mathbf{161.68}$&$\mathbf{10.29}$
&$\mathbf{18.11}$&$\mathbf{129.8680}$&$\mathbf{0.5682}$\\
127....&0.91&0.26&0.023&3.7257&-0.298&0.8381&4.45&169.93&169.97&170.11&170.42&170.11&12.62&21.33&130.5190&0.5917\\
128....&0.91&0.26&0.023&3.7267&-0.285&0.8470&5.05&167.28&167.34&167.47&167.78&167.47&12.01&20.49&130.5456&0.5861\\
129....&0.91&0.26&0.023&3.7277&-0.271&0.8564&5.65&164.54&164.61&164.73&165.05&164.73&11.39&19.62&130.5533&0.5805\\

\hline
 \end{tabular}\]}\\
 Note.---The above mean large separations were calculated averaging over n=10, 11, 22, ...,
 30. The mean small separations were averages over n=10, 11, 12, ...,
 30 at a fixed $l$ (as indicated).
 \end{table*}

\end{document}